# Title

**Full title:** Epileptic seizure forecasting with long short-term memory (LSTM) neural networks

**Short title:** Seizure forecasting with LSTMs

**Author list:** Daniel E. Payne[1,2,3], Jordan D. Chambers[1], Anthony Burkitt[1], Mark J. Cook[2,3], Levin Kuhlman[4], Dean R. Freestone[2] and David B. Grayden[1,3,5]

[1]Department of Biomedical Engineering, The University of Melbourne, Melbourne, VIC, Australia.

[2]Seer Medical, Melbourne, VIC, Australia.

[3]Department of Medicine, St Vincent's Hospital, The University of Melbourne, Melbourne, VIC, Australia.

[4]Department of Data Science and AI, Faculty of Information Technology, Monash University, Clayton, Australia

[5]Graeme Clark Institute for Biomedical Engineering, The University of Melbourne, Melbourne, VIC, Australia.

**Corresponding author email:** jordanc@unimelb.edu.au (JC)

**Author contributions:**

Conceptualization: DP, MC, LK, DF, DG

Data curation: DP, JC

Formal analysis: DP, JC, LK

Writing - original draft: DP, JC

Writing – review and editing: DP, JC, AB, MC, LK, DF, DG

# Abstract

**Objective:** Forecasting epileptic seizures can reduce uncertainty for patients and allow preventative actions. While many models can predict the occurrence of seizures from features of the EEG, few models incorporate changes in features over time. Long Short-Term Memory (LSTM) neural networks are a machine learning architecture that can display temporal dynamics due to the recurrent connections. In this paper, we used LSTMs to monitor changes in EEG features over time to improve the accuracy of seizure forecasts and to alter the time window of the forecast. **Methods:** Long-term intracranial EEG recordings from eight patients from the NeuroVista dataset were used. A Fourier transform of 1-minute segments of EEG was fed into a Convolutional Neural Network (CNN). The outputs from the CNN were input to three different LSTM models at different time intervals: 1 minute, 1 hour and 1 day. The LSTM model outputs were used to predict seizure onset within a time window. The prediction and start of the time window were separated by the same length of time as the window. Window sizes tested included 2, 4, 10, 20 and 40 minutes. **Results and Conclusion:** Our model forecast seizure onsets well above a random predictor. Compared to other models using the same dataset, our model performed better for some patients and worse for others. Monitoring the change in EEG features over time allowed our model to produce good results over a range of different window sizes, which is an improvement on previous models and raises the possibility of altering the forecast to meet individual patient needs. Furthermore, a window size of 40 minutes provides a potential intervention time of 40 minutes, which is the first time an intervention time of more than 5 minutes have been forecast using long-term EEG recordings.

## Introduction

Fifty million people, just under 1% of the world's population, have epilepsy. Typical epilepsy management is centred on the use of anti-epileptic drugs (AEDs), and surgical intervention for those patients who do not respond to AEDs. However, approximately 30% of people with epilepsy have seizures refractory to available medical and surgical treatment options. The uncertainty caused by the constant threat of seizures is the most significant factor impacting a person's quality of life with refractory epilepsy [1, 2]. Seizure forecasters aim to alleviate this uncertainty by advising when seizures will occur. In addition to reducing the uncertainty, seizure forecasters could allow for acute seizure preventions.

Most seizure forecasters focus on identifying and utilizing useful features extracted from electroencephalography (EEG). These features include measures of power [3, 4], synchronicity [5, 6] and entropy [7, 8], among others. Features are usually considered with no memory of previous feature values. For example, power could be measured over a 10-minute sample to predict seizure likelihood in the next 10 minutes. However, after that forecast, the power information is discarded and is not considered when making the next 10-minute prediction. This methodology assumes that the value of a feature is what is important without considering the importance of the change in a feature over time.

Cycles of hours to days can affect seizure likelihood [9, 10]. Evidence of these cycles can be found in EEG features and can be used to develop algorithms that outperform previous forecasters that used static features [3, 11, 12]. The dynamics of this system matters and, therefore, algorithms that aim to forecast seizures should attempt to account for these dynamics.

Long-term data must be used to capture and characterise long-term cycles. However, most EEG datasets are only 1-9 days in total recording time and so fall well short of the months to years required to capture long-term cycles. Datasets using seizure diaries are often long enough to capture these cycles, but they are less reliable and detailed than EEG; patients tend to underreport seizures when using a seizure diary [13]. The NeuroVista dataset [13], consisting of long-term intracranial EEG recordings of patients with localized epilepsies, has both the length and detail required to enable analysis of long-term cycles in EEG, which was not previously possible. However, the NeuroVista data includes 15 patients all with localized epilepsies, and so limits the generalization of the findings, though it provides an indication of what is possible.

Temporal factors have been shown to be effective in seizure forecasting [10, 14]. Cycles of seizure likelihood have been shown to occur over varying timescales, with 12-hour and 24-hour cycles being the most common, but cycles over a month in length have also been identified [9, 15]. Examining critical slowing over long timescales led to the most promising results with the NeuroVista dataset yet [12]. As effective as the critical slowing performance was, other factors may also impact seizure likelihood over this long timescale, potentially further improving forecasters. It is not known whether a deep neural network could capture information over a long timescale without programming features explicitly. The question arises then whether neural networks can effectively extract useful information of long-term cycles without hardcoded feature engineering or whether feature engineering is the better approach because of the limitations of the NeuroVista dataset.

The requirements for a good forecaster include specifying a seizure prediction horizon and a seizure onset period [16, 17]. The seizure prediction horizon defines the time between a forecast and the onset of a seizure, which provides the potential for intervention. Part of the reason why different studies use different seizure prediction horizons is because it is not known what the best seizure prediction horizon is for patients. Different interventions would require different seizure prediction horizons. For example, a system that provides electrical stimulation of the brain to alleviate a seizure could have an intervention period of less than a minute, whereas a system that requires the patient to take oral medication to alleviate a seizure would require an intervention period of more than 15 minutes. Furthermore, one of the most debilitating aspects of epilepsy is the uncertainty of when a seizure will happen. So, an advisory system that does not try to intervene and stop the seizure would still be very useful for patients. However, patients would still require and/or desire a seizure prediction horizon to prepare for the seizure. It is not clear what seizure prediction horizon is desired by patients and can vary between patients [16, 17]. There is also the potential for the desired seizure prediction horizon to change for the same patient when performing different activities. Therefore, a model that can alter the seizure prediction horizon would be very beneficial.

This study investigated the effectiveness of LSTMs for seizure forecasting with minimal feature engineering. Multiple LSTMs were trained to predict seizures over multiple timescales. Data was first fed into CNNs to identify features, and then the LSTMs identified long-term patterns in the CNN-learned features to forecast seizures. These models were used together to capture the level of detail needed while operating over very long timescales. By combining the pattern-learning power of machine learning with the proven effectiveness of forecasting over long timescales, it was hoped that generated seizure forecasts could improve upon previous attempts, taking one step closer to relief from uncertainty for people with epilepsy.

# Materials and Methods

## Datasets

Long-term intracranial EEG from the NeuroVista dataset was used [13]. We accessed patient data on the University of Melbourne servers and this dataset was generated on 4 July 2017. We did not have access to any information that could identify individual patients. Further information regarding the NeuroVista dataset can be found at www.epilepsyecosystem.org and in the original clinical trial [13]. Each EEG recording consisted of 16 electrodes sampled at 400 Hz for 1.5 years on average (range 1.0-2.1 years). Of the original 15 patients, eight were used in this study. Patients 2, 4, 5, 7, 12 and 14 were excluded as they were deemed to have too few seizures to train a neural network. Patient 3 was excluded due to a high amount of data loss during recording. Further information on the collection of the data can be found in the original NeuroVista publication [13].

The first 100 days of the recordings were excluded from each dataset due to the inconsistency of the recordings in this time period [18]. The remaining data was split into training and testing sets pseudo-prospectively with an 80:20 split. The first 80% of seizures was allocated as the training set and the remaining 20% as the testing set for each patient. The cut-off between training and testing sets was marked at the half-way point between the last training seizure and the first testing seizure, with all interictal samples before the cut-off being part of the training set and those after part of the testing set. Only lead seizures were used, where a lead seizure is defined as not having a seizure in the four hours prior to the seizure.

Each set was segmented into samples (periods of time) defined by the prediction window size. We tested prediction window sizes of 2, 4, 10, 20 and 40 minutes. Samples started from the beginning of the set (100 days from recording start and the cut-off for the training and testing sets, respectively). This resulted in samples that were not aligned to seizure start, which was necessary to replicate real-world conditions for a forecaster. Each sample was labelled as either interictal, ictal or preictal. Figure 1 illustrates the timing of samples, forecasts, and seizures.

Samples were labelled *ictal* if the seizure started within that sample. The following sample was also labelled ictal to account for seizures that continued into the next sample and for the postictal suppression period [14]. The samples that were two samples prior to the first ictal samples were labelled as *preictal* samples. This means the forecast, made at the end of the preictal period, applied to a seizure occurring twice the duration of the sample later, creating a seizure prediction horizon. All other samples were labelled as *interictal*. This labelling configuration meant that there was an intervention time of one prediction window and a seizure occurrence period of one prediction window, matching seizure forecasting requirements [16, 17].

## The model

### Overview

The overall model structure is shown in Figure 2. 1-minute segments of EEG were converted to the time-frequency domain (a spectrogram) by performing a Fourier transform for each electrode. Only frequencies up to 120 Hz were used. A window size of 1 second was used for the Fourier transform, which was shifted by 0.5 s thereby creating 120 data points along the time axis. The resulting 230,400 data points (16 electrodes x 120 Hz x 120 timesteps) were passed through a Convolutional Neural Network (CNN) to enable the use of high-frequency patters over longer timescales. The output of the CNN was 1 data point, simplifying the data for easier training of the LSTMs.

The outputs of the CNN were fed into three different LSTMs named Short, Medium and Long, which each captured information over different timescales. *Short* took CNN outputs every minute and made forecasts every prediction window or every minute. *Medium* took CNN outputs once per hour and made a forecast once per hour. *Long* took CNN outputs once per day and made a forecast once per day. LSTMs covering different timescales were compared to gain insight into which timescales may prove useful in the design of a seizure forecaster.

### CNN

The following architecture was the base version of the CNN. The design of the algorithm was iteratively improved as aspects of CNN design was tested. Details of these iterative improvements are shown in the Supporting Information. **Error! Reference source not found.Error! Reference source not found.**

The CNN (Figure 3) consisted of two convolutional layers with 16 neurons in the first layer and 32 neurons in the second layer. CNN kernels were 5-wide in the time dimension and 16-deep across the EEG channels. Convolution was not carried out across the EEG channel dimension. There was no dilation in the kernels for any layer. A Rectified Linear Unit (ReLU) was used as the activation function for the convolutional layers. After each convolutional layer, max-pooling was performed with a size and stride of 4; this outputs the maximum value of every four values. Four was chosen instead of the typical two to reduce the size of the time dimension to a larger extent as the input size was 23,960 values per channel. After each max-pooling layer, batch normalization was performed.

After the convolutional layers, a dense layer with 16 neurons was added, again using ReLU as the activation function, followed by batch normalization. A final single-neuron dense layer was added to produce the single value forecast. The output of the last neuron was passed through a sigmoid function to produce a forecast between 0 and 1: 1 represents data labelled as pre-ictal and 0 represents data labelled as inter-ictal.

The training dataset for the CNN model used down-sampling of the inter-ictal samples to produce a balanced dataset. Interictal samples were chosen from samples 5 hours or more from a seizure. They were also limited to being at least 2 hours from another interictal sample so that interictal samples were spread more evenly through the data.

### LSTM

The LSTM architecture (Figure 3) was the same for the Short, Medium and Long models. There was a single LSTM layer, followed by a Dense layer to create a single forecast. The CNN outputs were fed into the LSTM layer. The different LSTM models received data at different time scales from the CNN model. The Short LSTM model received input every minute, with a total of 60 samples taken from the previous 90 minutes (the closest 60 samples chosen to the seizure were chosen to allow for data dropouts). The Medium LSTM model received input every hour, with a total of 24 samples over the previous day (to account for data dropouts, the closest 1-minute sample to the desired time was used, provided it was within 30 minutes of the desired time). The Long LSTM model received input every day, with a total of 30 samples over the previous 30 days (to account for data dropouts, the closest 1-minute sample to the desired time was used, provided it was within 12 hours of the desired time). Since the LSTM models had different numbers of inputs, the numbers of neurons in the LSTM layers were chosen accordingly. The Short LSTM had 128 neurons, the Medium LSTM had 16 neurons and the Long LSTM had 32 neurons. Each of the LSTM models had the same hyperparameters: a sigmoid activation, a sigmoid recurrent activation and 0.25 recurrent dropout.

When data dropouts prevented the correct number of samples being generated for each of the LSTM models, that data was removed from the training and test datasets. Therefore, to increase the dataset sizes, the training dataset for the LSTM models used up-sampling of the pre-ictal samples to produce a balanced dataset. When up-sampling, noise of ±5% was added to the values using a uniform random distribution.

For both the CNN and LSTM models, Adam [19] was used as the optimization function with a learning rate of 0.001.

### Combination model

Initial results for the different LSTM models showed significant variability between different patients and different prediction windows. There was no consistency to determine the best model nor best prediction window. Therefore, we created a model that could combine the outputs of all other models into a single forecast. This model (Figure 4) comprised a dense (fully connected) layer, a dropout layer and a final dense layer. The model received input from the 1-minute CNN model, Short LSTM model, Medium LSTM model, Long LSTM model and information describing the time of day. Information describing the time of day included hour of the day, day of the week and day of the month. Each of these values was presented as an absolute value and a minimum distance from the start (to indicate the cyclic nature of time) at a resolution of 1-minute. For both the training and test datasets, samples were created whenever there was an output value for the 1-minute CNN model. If data was missing due to data dropouts from one or more of the LSTM models, a value of zero was used as the input.

Since the combination model also introduced time of day information, we also tested a model that only included time of day information, referred to as the Machine Learning Time of Day Model (ML-TOD). This model was identical to the combination model and was trained and tested on identical datasets, but did not receive any inputs from the CNN model or the LSTM models.

### Statistical comparisons

To compare the results of different models, we used the receiver operator characteristic (ROC) curve, which calculates the performance of a classification model at all possible values through the relationship between the true positive rate and false positive rate. The true positive rate is defined as the true positives as a proportion of the true positives and false negatives. The false positive rate is defined as the false positives as a proportion of the false positives and true negatives. The area under the curve (AUC) measures the area underneath the entire ROC curve. To compare AUCs, we calculated the confidence intervals using the Hanley and McNeil method [20]. This method uses Wilcoxon statistic and takes into account the correlation between AUCs due to the paired nature of the data when using the same sample of patients, which increases the statistical power.

For a two-class problem, a random predictor will have an AUC of 0.5. While we did not estimate the confidence intervals of a random predictor (as it will just depend on the nature of the random predictor), we assumed the results were significantly different to a random predictor when the confidence intervals of our models AUCs did not overlap with a value of 0.5. When comparing two values with confidence intervals, we assumed a significant difference only when there was no overlap between the two confidence intervals.

## Results

We created a seizure forecaster that converted 1 minute of EEG recordings into a spectrogram that fed into a CNN model that was trained to make a seizure forecast. The output of the CNN model was then fed into three LSTM models that looked at how the seizure forecasts varied over different time periods. The final seizure forecast involved combining the output of the CNN model, three LSTM models and time of day information.

These models contained multiple hyperparameters. Hyperparameter exploration and optimisation is described in the supplementary information (S1).

### Patient Forecasts

Figure 5 shows the AUC results for all models and all patients. There are six models in total. The 1 minute model takes 1 minute of EEG recordings, converts it to a spectrogram and then the CNN model reduces it down to a single value. There are three LSTM models: short, medium and long. The short model takes 60 inputs from the CNN model over the previous 60 minutes. The medium model takes 24 inputs from the CNN model over the previous 24 hours, with a 1 minute sample selected per hour. The long model takes 30 inputs from the CNN model over the previous 30 days, with a 1 minute sample selected per day. The combo model receives ten inputs: one from the 1 minute CNN model, three from the three LSTM models and six inputs relating to the time of day. Finally, the Machine Learning Time Of Day (ML-TOD) model has the same model structure as the combo model, but only receives six inputs relating to the time of day and does not receive any information from the EEG recordings. All six models were trained and tested on different prediction windows sizes (2 minutes, 4 minutes, 10 minutes, 20 minutes and 40 minutes). The performance of each model and for each prediction window was compared to a random predictor (0.5 AUC for 2-label data) and a simple hour of day predictor [10, 11].

### Patient 1

For Patient 1 (Figure 5A), 28 out of 30 results were significantly better than random prediction and a simple hour of day predictor. It was only the short model at small prediction windows (2 minutes and 4 minutes) that did not perform well. The short model performed well at the larger prediction windows and produced the best performance of all models for the 20 minutes prediction window. The combination model was the most consistent overall, often producing the second-best performance for each prediction window. For Patient 1, the best results were observed for the 4 minutes prediction window.

### Patient 6

For Patient 6 (Figure 5B), 24 out of 30 results were significantly better than random prediction and a simple hour of day predictor. Four out of the six results that were not significantly different to random prediction occurred in the 40-minute prediction window, which indicates the EEG signals picked up by these models to predict a seizure occurred less than 40 minutes before the seizure. Indeed, there was a trend of decreased performance across all models when the prediction window was increased to more than 4 minutes. Furthermore, for the larger prediction windows, the time of day predictor performed better than the other models. Again, the combination model was the most consistent performer, producing the best result at the 4-minute prediction window and closely following the time of day predictor when the EEG-only models performed weaker at the larger prediction windows.

### Patient 8

For Patient 8 (Figure 5C), 24 out of 30 results were significantly better than random prediction. However, the hour of day predictor produced an excellent result of 0.73 AUC. Therefore, none of the results were significantly better than the hour of day predictor, but only five results were significantly less. All five results that were less came from the long LSTM model. For Patient 8, the long LSTM model struggled to train and produce useful predictions. Nearly all predictions from this model were solely pre-ictal or inter-ictal, rather than spread between the two options. As a result, the AUC could not be calculated and so results are not shown. It is noted that Patient 8 had a large number of lead seizures (0.85 per day) and a large number of total seizures (Supplementary Table S2) in both the training and test data. The large number of seizures can cause trouble with the long LSTM model as data is excluded for 4 hours after a seizure.

### Patient 9

For Patient 9 (Figure 5D), 24 out of 30 results were significantly better than random prediction. However, the hour of day predictor produced a result of 0.67 AUC. Therefore, only nine of the results were significantly better than the hour of day predictor. The combination model was always significantly better than the hour of day predictor. The ML TOD model also produced three results significantly better than the hour of day predictor and was never statistically different to the combination model, indicating that the combination model was largely using time of day information. The other model to produce a result significantly better than the hour of day predictor was the long LSTM model at a prediction window of 40 minutes. Similar to Patient 8, due to the large number of seizures for Patient 9, the long LSTM model could not make meaningful predictions with the smaller prediction windows.

### Patient 10

For Patient 10 (Figure 5E), 24 out of 30 results were significantly better than random prediction. Similar to Patients 8 and 9, the hour of day predictor produced a high result of 0.66 AUC. This meant that none of the results were statistically better than the hour of day predictor, but only five out of 30 results were significantly less. Similar to Patients 8 and 9, Patient 10 had a large number of lead seizures. Compared to Patients 8 and 9, the ML-TOD model did not perform as well for Patient 10, producing overall lower results. The combination model produced the most consistent results, closely followed by the time of day model. The medium LSTM produced the best result for the 2-minute prediction window and produced good results for the other prediction windows, whereas the short and long LSTM models were inconsistent across the different prediction windows.

### Patient 11

For Patient 11 (Figure 5F), all results were significantly better than random prediction and the hour of day predictor. The 1-min CNN model produced very good results for all prediction windows (0.68-0.84 AUC). Consequently, the short and long LSTM models also produced very good results and significantly improved the on results of the 1-min CNN model for 3 out of the 5 prediction windows. However, the medium LSTM model did not perform well overall and, for two prediction windows, was significantly less than the 1-min CNN model. The combination model closely followed the performance of the short and long LSTM models. The ML-TOD model performed well, but performed significantly lower than the short LSTM, long LSTM and combination models for three out of five prediction windows.

### Patient 13

For Patient 13 (Figure 5G), 29 out of 30 results were significantly better than random prediction and 27 out of 30 results were significantly better than the hour of day predictor. Despite the results being statistically significant, the overall performance was lower than other patients. The 1-min CNN model produced the best results at 4-minute and 2-minute prediction windows. All three LSTM models closely followed the performance of the 1-min CNN model, only producing a significantly better results for the short LSTM at a 20-minute prediction window and for the medium LSTM model at a 40-minute prediction window. The combination model produced the most consistent results over the different prediction windows. For three out of the five prediction windows, the combination model was significantly better than the ML-TOD model and, for two out of five prediction windows, the combination model was significantly better than the 1-minute CNN model. Once again, this indicates the combination model picked up the most useful information from the other models. Furthermore, the combination model was significantly better than both the 1-minute CNN model and the ML-TOD model for the 2-minute prediction window, which indicates there was a correlation between the three LSTM models (which were not significantly better than 1-minute CNN model or the ML TOD model) that the combination model was utilising to produce a significant increase in performance.

### Patient 15

For Patient 15 (Figure 5H), 22 out of 30 results were significantly better than random prediction and 21 out of 30 results were significantly better than the hour of day predictor. It was the 2-minute and 4-minute predictions windows where the models struggled to produce statistically significantly results, indicating the biomarkers useful to forecast seizures for Patient 15 were appearing 20 to 80 minutes before the seizure. The ML TOD model and the combination model were the most consistent performers for Patient 15. While the three LSTM models consistently performed better than the 1-minute CNN model, only on two occasions was this increase statistically significant, both of which occurred with the larger prediction windows.

### Overall model performance

When looking at the average performance across all eight patients (Figure 6), the combination model produced the best results for all prediction windows (0.72-0.75 AUC). The next best on average was the ML-TOD model (0.68-0.70 AUC), followed by the medium LSTM model (0.64-0.68 AUC), the short LSTM model (0.62-0.69), long LSTM model (0.57-0.72) and the 1-minute CNN model (0.60-0.65 AUC).

When comparing the performance of the different prediction windows, the average across eight patients did not show any trend or obvious difference (Figure 6). This is to be expected because the results for each patient showed different preferences for the different prediction windows. The spread of best performance across the different prediction windows was very even, with two patients producing their personal best performance for prediction windows of 2-minutes, 4-minutes, 10-minutes and 40-minutes. It was only the 20-minutes prediction window that did not have a best performance for any patients, but it often produced similar results to either 10-minutes prediction window or the 40-minutes prediction window.

### Comparisons to previously published models

To compare our results with previously published models, we used the forecasts from the combination model as this produced the most consistent results across all patients and incorporated the values of all other models. While there are many publications analysing EEG recordings to forecast seizures, the NeuroVista dataset is unique because it is long-term recordings from patients living their normal lives as compared to short-term recordings in clinical settings. Therefore, we limited the comparison to other studies that have used the NeuroVista dataset. Most of the previous NeuroVista studies have reported findings in terms of sensitivity and selectivity (often reported as Time in High), where sensitivity is the proportion of seizures accurately forecast and selectivity is the proportion of time the model is forecasting a seizure to occur within the seizure prediction window. Therefore, the forecasts of this current model were converted to sensitivity and selectivity, taking a value of 0.5 as the threshold for forecasting the occurrence of a seizure.

Table S2 shows the results in terms of sensitivity and selectivity for the combination model on all patients and all prediction windows. Table S2 demonstrates the prediction window can have a significant effect on both the sensitivity and selectivity for the same patient and same model. Therefore, a direct comparison to previously published models would be best performed with the same prediction window and same seizure prediction horizon. Unfortunately, almost every previously published model uses a different prediction window and prediction horizon. As this current work has the improvement of being able to vary the prediction window, we chose the prediction window that was the closest match to the previously published model. A direct comparison between sensitivity and selectivity is often difficult as it is a snapshot of the AUC at one threshold and both numbers are dependent on each other. To make a direct comparison with previous results, we produced the sensitivity and selectivity for a thousand different thresholds and selected the thresholds that best matched the sensitivity or selectivity for the previous results. When one of the values is matched, a direct comparison between the other values can readily be made.

### Comparison to the critical slowing model

The critical slowing model [12] measures the half-width of the autocorrelation of the EEG signal to look for critical slowing down that may predict a state change in the underlying physiology. This biomarker was found to show an increase in half-width between 0.5-3 minutes before the onset of a seizure. The autocorrelation is combined with other features and is input to a model that incorporates long and short term seizure cycles. The final prediction used a prediction horizon of 2-4 minutes, so we choose the 2-minute combination model to compare the results.

Table 1 compares the critical slowing model with the 2-minute combination model. The critical slowing model produces better results for all 8 patients. For 4 patients, the 2-minute combination model is approximately 10-20% below the critical slowing model, but for the remaining patients the difference is very large.

| | Critical slowing model | | 2-min combination | | 2-min combination with matched sensitivity | | 2-min combination with matched selectivity | |
|---|---|---|---|---|---|---|---|---|
| | Sensitivity | Selectivity | Sensitivity | Selectivity | Sensitivity | Selectivity | Sensitivity | Selectivity |
| Patient 1 | 0.83 | 0.08 | 0.50 | 0.14 | 0.84 | 0.30 | 0.31 | 0.08 |
| Patient 6 | 0.66 | 0.03 | 0.40 | 0.21 | 0.66 | 0.41 | 0.23 | 0.03 |
| Patient 8 | 0.64 | 0.23 | 0.72 | 0.36 | 0.64 | 0.27 | 0.58 | 0.23 |
| Patient 9 | 0.85 | 0.16 | 0.88 | 0.33 | 0.85 | 0.32 | 0.61 | 0.16 |
| Patient 10 | 0.78 | 0.24 | 0.41 | 0.23 | 0.78 | 0.72 | 0.43 | 0.24 |
| Patient 11 | 0.86 | 0.16 | 0.73 | 0.19 | 0.86 | 0.28 | 0.68 | 0.16 |
| Patient 13 | 0.64 | 0.14 | 0.47 | 0.18 | 0.64 | 0.31 | 0.42 | 0.14 |
| Patient 15 | 0.87 | 0.0007 | 0.44 | 0.25 | 0.87 | 0.58 | 0 | 0.007 |

*Table 1:* A comparison of the critical slow model with the 2-minute prediction window and combination model. 2-min combination with matched sensitivity columns allows a direct comparison between the selectivity of the critical slowing model. 2-min combination with matched selectivity columns allows a direct comparison between the sensitivity of the critical slowing model.

### Comparison to the deep CNN model

The deep CNN model [3] generated a spectrogram of the EEG signal and used a CNN to perform image recognition on the spectrograms plus an image incorporating the time of day. This model defined the pre-ictal period as 16 minutes to 1 minute before the seizure onset. While this does not directly compare to any of the prediction windows used in this study, the best comparison is a prediction window of 10-minutes, which defines the pre-ictal period as 20 minutes to 10 minutes before seizure onset.

Table 2 compares the Deep CNN model with the 10-minute combination model. The 10-minute combination model performed 5-10% better for two patients (Patient 1 and Patient 15), both models were almost identical for one patient (Patient 8), the Deep CNN model was less than 5% better for two patients (Patient 9 and Patient 10) and was greater than 15% better for two patients (Patient 11 and Patient 13).

| | Deep CNN model | | 10-min Combination model | | 10-min combination model with matched sensitivity | | 10-min combination model with matched selectivity | |
|---|---|---|---|---|---|---|---|---|
| | Sensitivity | Selectivity | Sensitivity | Selectivity | Sensitivity | Selectivity | Sensitivity | Selectivity |
| Patient 1 | 0.65 | 0.21 | 0.40 | 0.11 | **0.65** | **0.18** | **0.71** | **0.21** |
| Patient 8 | 0.77 | 0.32 | 0.74 | 0.31 | 0.77 | 0.33 | 0.76 | 0.32 |
| Patient 9 | 0.83 | 0.43 | 0.64 | 0.23 | 0.83 | 0.46 | 0.81 | 0.43 |
| Patient 10 | 0.68 | 0.32 | 0.54 | 0.23 | 0.68 | 0.38 | 0.65 | 0.32 |
| Patient 11 | 0.78 | 0.18 | 0.76 | 0.24 | 0.78 | 0.25 | 0.64 | 0.18 |
| Patient 13 | 0.70 | 0.21 | 0.43 | 0.28 | 0.70 | 0.57 | 0.32 | 0.21 |
| Patient 15 | 0.59 | 0.37 | 0.36 | 0.22 | **0.59** | **0.30** | **0.68** | **0.37** |

*Table 2:* A comparison of the deep CNN model with the 10-minute prediction window and combination model. 10-min combination with matched sensitivity columns allows a direct comparison between the selectivity of the Deep CNN model.

10-min combination with matched selectivity columns allows a direct comparison between the sensitivity of the Deep CNN model. Values highlighted in bold indicate patients where our algorithm performed better than the deep CNN model.

## Comparison to the Kaggle competition best models

A Kaggle competition "Melbourne University AES/MathWorks/NIH Seizure Prediction" was run using data of three patients from the NeuroVista dataset [11]. There were 646 competition participants, 478 teams and over 10 000 algorithms submitted in the competition. The winning team's submission involved a combination of an ensemble of different models. In the competition, the pre-ictal period as 65 minutes to 5 minutes before the seizure onset. While this does not directly compare to any of the prediction windows used in this study, the best comparison is a prediction window of 20 minutes, which defines the pre-ictal period as 40 minutes to 20 minutes before seizure onset.

Table 3 compares the results from the best performing model of the Kaggle competition to the 20-minute combination model. The best of the Kaggle competition performed better for both patients, less than 5% better for Patient 9 and less than 10% better for Patient 10. The third patient is not compared as this one was not included in the current study.

|  | Kaggle | | 10-min Combination model | | 10-min combination model with match sensitivity | | 10-min combination model with match selectivity | |
|---|---|---|---|---|---|---|---|---|
|  | Sensitivity | Selectivity | Sensitivity | Selectivity | Sensitivity | Selectivity | Sensitivity | Selectivity |
| Patient 9 | 0.52 | 0.11 | 0.73 | 0.29 | 0.52 | 0.13 | 0.46 | 0.11 |
| Patient 10 | 0.53 | 0.17 | 0.70 | 0.43 | 0.53 | 0.25 | 0.42 | 0.17 |

*Table 3:* A comparison of the best models from the Kaggle competition with the 10-minute prediction window and the combination model. 10-min combination with matched sensitivity columns allows a direct comparison between the selectivity of the Kaggle competition winners. 10-min combination with matched selectivity columns allows a direct comparison between the sensitivity of the Kaggle competition winners.

## Comparison to the original NeuroVista clinical trial

The original NeuroVista clinical trial [13] performed a spectral analysis of the EEG recordings to create feature vectors that were then used in a classifier that was a combination of a decision tree and nearest neighbour classifier. This model made predictions of high, moderate or low seizure risk. High was considered a correct forecast if a seizure occurred in the next 5 minutes. A seizure onset period of 5 minutes is best matched by a prediction window of 4-minutes in our model.

Table 4 compares the performance of the original NeuroVista clinical trial with the 4-minute combination model. The 4-minute combination model performs more than 20% better for 2 patients (Patient 9 and Patient 11), approximately 10% better for 3 patients (Patient 1, Patient 8 and Patient 13), approximately 10% worse for Patient 10 and more than 30% worse for Patient 15.

|  | Original Trial | | 4-min Combination model | | 4-min Combination model with matched sensitivity | | 4-min combination model with matched selectivity | |
|---|---|---|---|---|---|---|---|---|
|  | Sensitivity | Selectivity | Sensitivity | Selectivity | Sensitivity | Selectivity | Sensitivity | Selectivity |
| Patient 1 | 0.77 | 0.27 | 0.60 | 0.14 | **0.77** | **0.15** | **0.80** | **0.27** |
| Patient 8 | 0.62 | 0.28 | 0.80 | 0.39 | **0.62** | **0.20** | **0.72** | **0.28** |
| Patient 9 | 0.17 | 0.11 | 0.87 | 0.36 | **0.17** | **0.02** | **0.49** | **0.11** |
| Patient 10 | 0.51 | 0.17 | 0.62 | 0.39 | 0.51 | 0.26 | 0.42 | 0.17 |
| Patient 11 | 0.39 | 0.15 | 0.72 | 0.20 | **0.39** | **0.05** | **0.65** | **0.15** |
| Patient 13 | 0.50 | 0.28 | 0.62 | 0.31 | **0.50** | **0.21** | **0.58** | **0.28** |
| Patient 15 | 0.71 | 0.21 | 0.38 | 0.28 | 0.71 | 0.56 | 0.32 | 0.21 |

*Table 4:* A comparison of the results from the original clinical trial with the 4-minute prediction window and combination model. 4-min combination with matched sensitivity columns allows a direct comparison between the selectivity of the original clinical trial. 4-min combination with matched selectivity columns allows a direct comparison between the sensitivity of the original clinical trial. Values highlighted in bold indicate patients where our algorithm performed better than the original clinical trial.

### Comparison to the circadian logistic regression model

The circadian logistic regression model [10] extract 80 signal features from the EGG recordings relating to the energy in different frequency bands and performed logistic regression to a spectral analysis of the EEG recordings to create to feature vectors that were then used in a classifier that was a combination of a decision tree and nearest neighbour. The model produces a forecast that a seizure would occur in the next 30 minutes. While this does not directly compare to any of the prediction windows used in this study, the best comparison is a prediction window of 20 minutes, which defines the pre-ictal period as 40 minutes to 20 minutes before seizure onset.

Table 5 compares the performance of the circadian logistic regression model with the 20-minute combination model. The 20-minute combination model performs does not perform better for any patients. However, it produces the same performance for Patient 9 and is less than 10% worst for 4 patients (Patient 1, Patient 8, Patient 10 and Patient 15).

| | Circadian logistic regression | | 20-min Combination model | | 20-min Combination model with matched sensitivity | | 20-min combination model with matched selectivity | |
|---|---|---|---|---|---|---|---|---|
| | Sensitivity | Selectivity | Sensitivity | Selectivity | Sensitivity | Selectivity | Sensitivity | Selectivity |
| Patient 1 | 0.61 | 0.27 | 0.54 | 0.20 | 0.61 | 0.33 | 0.56 | 0.27 |
| Patient 8 | 0.76 | 0.28 | 0.75 | 0.35 | 0.76 | 0.35 | 0.68 | 0.28 |
| Patient 9 | 0.45 | 0.11 | 0.77 | 0.29 | 0.45 | 0.11 | 0.46 | 0.11 |
| Patient 10 | 0.52 | 0.17 | 0.67 | 0.37 | 0.52 | 0.25 | 0.42 | 0.17 |
| Patient 11 | 0.58 | 0.15 | 0.69 | 0.28 | 0.58 | 0.23 | 0.35 | 0.15 |
| Patient 13 | 0.76 | 0.28 | 0.57 | 0.30 | 0.76 | 0.56 | 0.57 | 0.28 |
| Patient 15 | 0.60 | 0.21 | 0.53 | 0.32 | 0.60 | 0.25 | 0.54 | 0.21 |

*Table 5:* A comparison of the results from the circadian logistic regression model with the 20-minute prediction window and combination model. 20-min combination with matched sensitivity columns allows a direct comparison between the selectivity of the original clinical trial. 20-min combination with matched selectivity columns allows a direct comparison between the sensitivity of the original clinical trial.

### Discussion

Accurate seizure forecasting has the potential to significantly improve an epilepsy patient's life by removing uncertainty and providing an opportunity for intervention or alleviation. Previously published algorithms can forecast seizures from features of the EEG, but few incorporate changes in features over time. We developed a model that incorporates changes in EEG features over time by using Long Short-Term Memory neural networks (LSTMs), a machine learning architecture that can display temporal dynamics. We used long-term intracranial EEG recordings from eight patients from the NeuroVista dataset [13]. A Fourier transform of 1-minute segments of EEG was used an input into a CNN mode. The output from the CNN model was played into three different LSTM models. The outputs from the CNN model and three LSTM models were then combined with time of day information through a dense, or full-connected, classifier. The performance of this model was competitive with the best models using the same dataset. This model structure and incorporating the changes in EEG features over time allowed for an alternation of the prediction window for the seizure forecast, which is an improvement on previous models.

## Forecasting seizures

A good seizure forecaster should contain a prediction horizon (a period of time between a forecast and onset of a seizure) to allow for potential interventions [16, 17]. This period is likely to be of different lengths of time for different interventions and different patients. We have demonstrated that our model can alter the seizure prediction horizon and still produce forecasts that are well above chance. This is the first study to demonstrate this while using long-term iEEG recordings. The fact that our model can easily change the prediction window opens the possibility that the advisory system could use patient-specific prediction windows and/or patients could change the prediction window according to their currents needs. Furthermore, our prediction window of 40 minutes provides a seizure prediction horizon of 40 minutes, which is the first study to demonstrate it is possible to forecast seizures from long-term iEEG recordings with an intervention time of more than 5 minutes. This larger seizure prediction horizon opens the possibility of different types of interventions.

The main difference between our model and previous seizure forecasters is that our model looked at the change of EEG features over time. We did this by using the output of the 1-minute CNN model as input into three different LSTM models. The performance of the LSTM models compared to the 1-minute CNN model was highly variable depending on the patient and prediction window chosen. Overall, 17% of the LSTM models were significantly better than the 1-minute CNN model, whereas only 3% of the LSTM models were significantly worse than the 1-minute CNN model. This indicates the change of EEG features over time can improve seizure forecasts. Furthermore, 68% of the combination models were significantly better than the 1-minute CNN model, whereas only 38% of the machine learning time of day predictor were better than the 1-minute CNN model (and 5% worse than the 1-minute CNN model). Not only does this again demonstrate changes of EEG features over time can improve seizure forecasts, but it also shows that a simple classifier can identify which data to focus on for that patient and prediction window.

Demonstrating that the change of EEG features over time can improve seizure forecasts is important because the 1-minute CNN model used in this study is relatively simple compared to the deep CNN model [3] and other previous models. Therefore, improving the 1-minute CNN model would expect to improve the forecasting performance. Similarly, the performance of our machine learning time of day predictor was significantly better than chance and significantly better than a simple hour of day predictor, but the performance probably was not as high as it could be given retrospective studies looking at the same information [15]. This is most likely due to our model being developed and trained to be used with the 1-minute CNN model and LSTM models. As a result, it only received information regarding the time of day when there were EEG recordings that could be used as input for the 1-minute CNN model. It did not receive information when there was a dropout in the EEG recordings, for 4 hours after a seizure, and only received information regarding the lead seizure (and not multiple seizures). Improving this model is also likely to improve the seizure forecasting performance.

## Comparisons to previously published models

To compare the performance of this current model with previously published models, we only considered models that used the same dataset. Short-term studies and studies recording EEG in clinical settings can produce time correlated data [21], allowing forecasting models to accurately predict seizures on those datasets, but the results do not translate to other datasets or prospective settings. Long-term recordings can more easily avoid this issue because they generate enough data to produce a training and testing datasets that are separate in time, thereby avoiding time correlated data. Furthermore, collecting data while patients are living their normal lives produces more variation in the EEG recordings, which appears to reduce time correlated data when looking at short periods of time in the long-term recordings (Chambers and West, unpublished).

Even when comparing models that use the same dataset, it is difficult to make direct comparisons because each investigation used different methods to describe the data. For example, the deep CNN paper labelled data as pre-ictal when 16-1 minutes before a seizure [3], whereas the Kaggle competition labelled data as pre-ictal when 65-5 minutes before a seizure [11]. Such differences can have a large effect on the results because it changes the amount of data available to optimise a model and can change the EEG features or biomarkers required to make useful forecasts. Our current results demonstrate this through changes in performance while varying the prediction window for different models and different patients.

While our model was successfully able to change the prediction window, our systematic method to alter the seizure prediction horizon and seizure onset period meant that none of the choices were an exact match to the values used in previous studies. Therefore, we choose the closest match from our five different prediction windows to each of the previous studies. The performance of our model was well below the critical slowing model [12]. The performance of the critical slowing model is state-of-the-art and significantly better than all other models published to date. However, the increased performance of the critical slowing model appears to be due to the autocorrelation feature often appearing 2 minutes or less before the seizure (a lot of other models exclude this information through their choice of seizure prediction horizon) and carefully curated data which requires clinician input initially. A better comparison is to the deep CNN model [3], which only received the same information: EEG data and time of day. Compared to the deep CNN model, our model performed better for two patients, very similar for three patients and worse for two patients. Similarly, comparisons with other models [10, 11, 13] have some patients where performance is better, some patients where performance is worse and some where it is very similar. Overall, this indicates the performance of our model is competitive with previously published models. It would be expected that the CNN model would perform better than our current 1-min CNN model because it has a larger and deeper architecture with more parameters. We kept our 1-min CNN model relatively simple to reduce computational times, so it would be expected that using a larger architecture would reproduce the same performance as the CNN model. Since our model performed better for two patients and very similar for three patients, this indicates that using LSTMs to interpret the change in features over time is an improvement.

## Conclusion

We developed a model with minimal feature extraction that can forecast epileptic seizures while varying the seizure prediction horizon. Altering the seizure prediction horizon is a significant improvement because it allows for patient-specific warnings and opens the possibility of new or different interventions. Our work has demonstrated that incorporating changes in EEG features over time can improve seizure forecasting.

# Figures

## Figure 1
A schematic diagram indicating how the EEG data is labelled inter-ictal or pre-ictal. SPH stands for seizure prediction horizon. SOP stands for seizure onset period. *w* represents the prediction window. We used prediction windows of 2 minutes, 4 minutes, 10 minutes, 20 minutes and 40 minutes. Data was labelled pre-ictal when located between 1 and 2 prediction window sizes before the SOP. Data was labelled inter-ictal when located more than 3 prediction window sizes before the SOP.

## Figure 2
A schematic diagram showing the overall structure of the model. Intracranial EEG (iEEG) recordings from 16 electrodes are taken from patients. iEEG recordings are converted to the frequency domain by taking a spectrogram of 1 minute of data. Spectrograms are fed into convolutional neural network (CNN), which is trained to classify the image as inter-ictal or pre-ictal. The output of the CNN model at different time points is fed into a long-short term memory (LSTM) neural network to make the final forecast.

## Figure 3
A schematic diagram showing the structure of the CNN model and the LSTM model. The CNN model received a spectrogram as the input data. It comprised of convolutional layer, max pooling layer, convolutional layer, max pooling layer and two dense (or fully connected) layers. The output of the 1-min CNN model was used as the input into the LSTM model.

## Figure 4
A schematic diagram showing the structure of the combination model. The combination model received input from the 1-min CNN model, Short LSTM model, Medium LSTM model, Long LSTM model and time of day information. The model comprised of dense (or fully connected) layer, a dropout layer, and a final dense (or fully connected) layer.

## Figure 5
Indicates the area under the curve (AUC) for 6 different models and 5 different prediction windows. Panels A-D display the same information for different patients. Error bars indicate the Hanley and McNeil confidence intervals for comparisons of AUCs. The solid black line indicates the AUC for a random predictor. The solid red line indicates the AUC for a simple cyclic predictor using the hour of day and the dashed red lines indicate the Hanley and McNeil confidence intervals of the cyclic predictor.

## Figure 6
Indicates the average area under the curve (AUC) for 6 different models and 5 different prediction windows. The average was taken from all 8 patients across the different models and prediction windows. In cases when an AUC was not calculated due to missing data, these values were excluded from the average. Error bars in the 90% confidence interval.

Figure 1

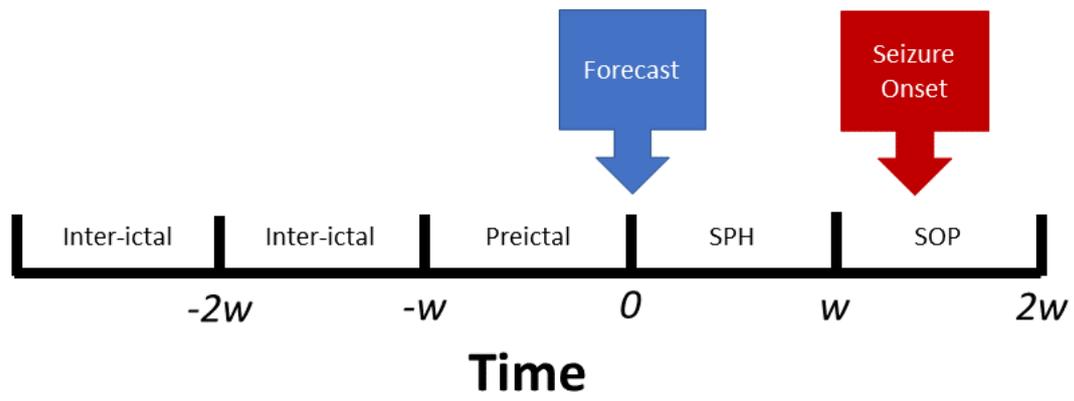

Figure 2

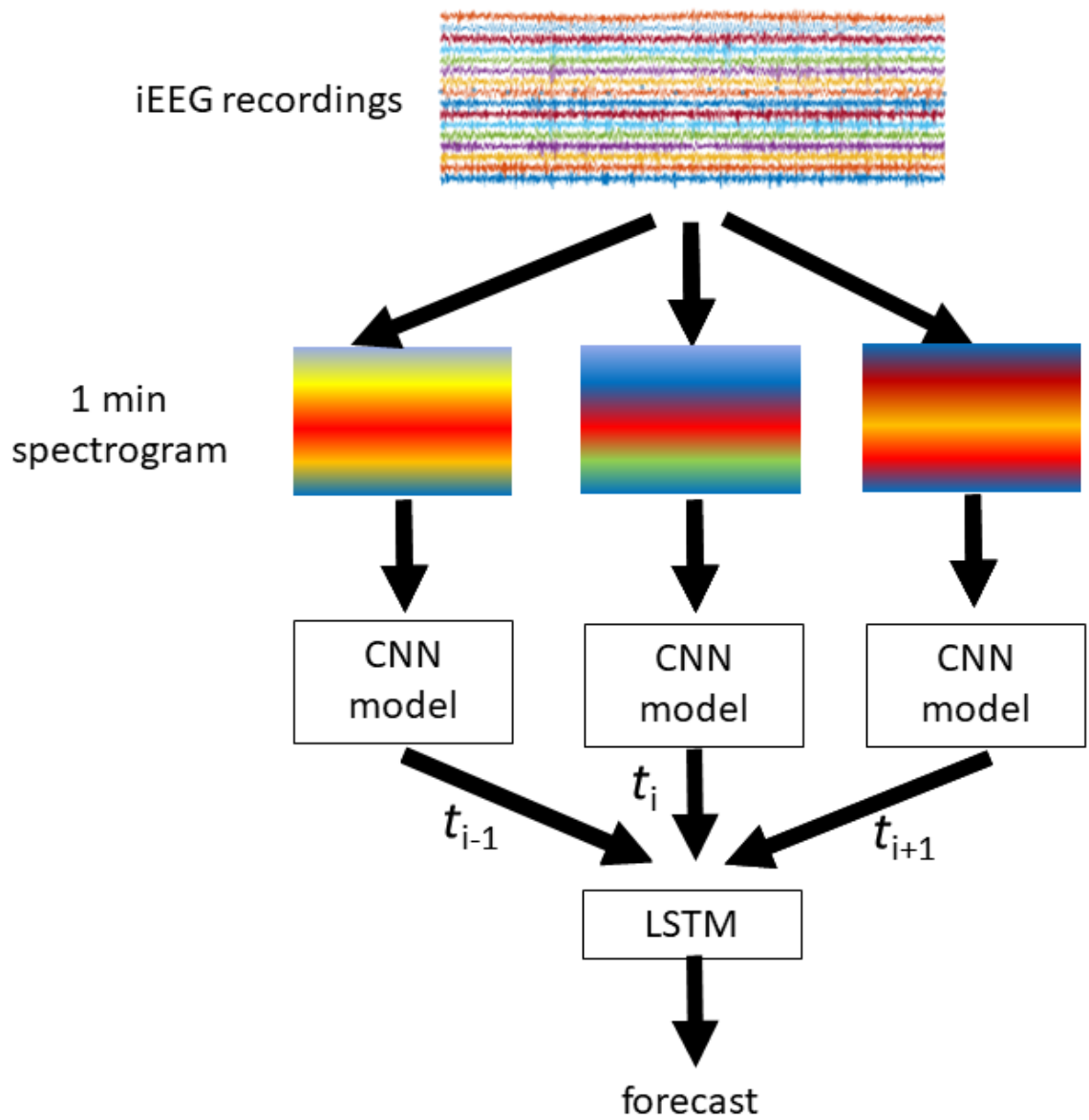

# Figure 3

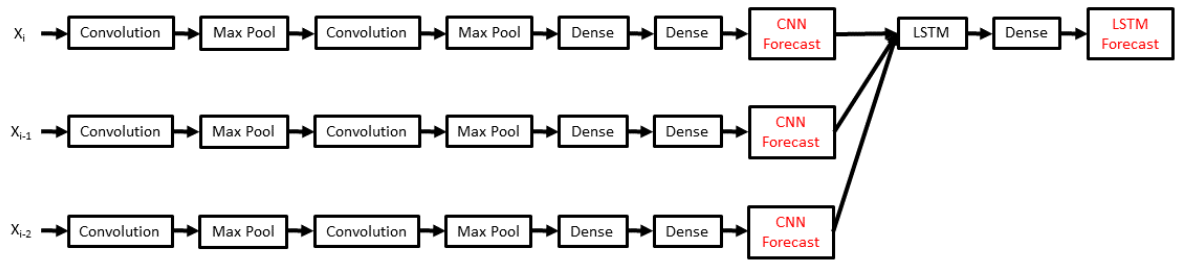

Figure 4

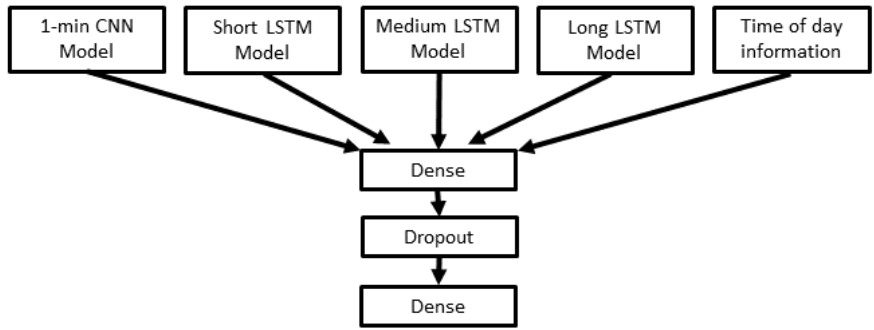

Figure 5

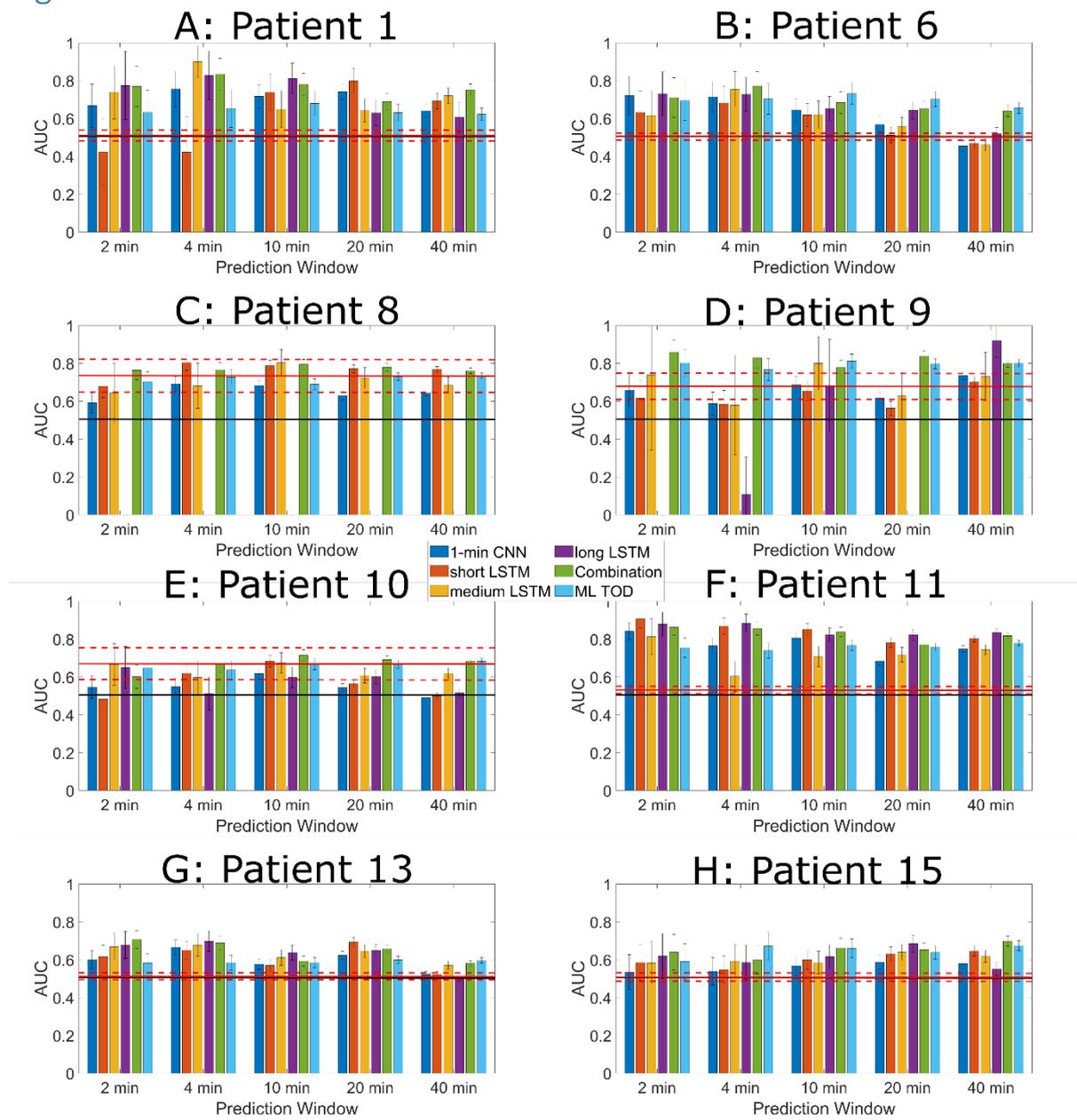

Figure 6

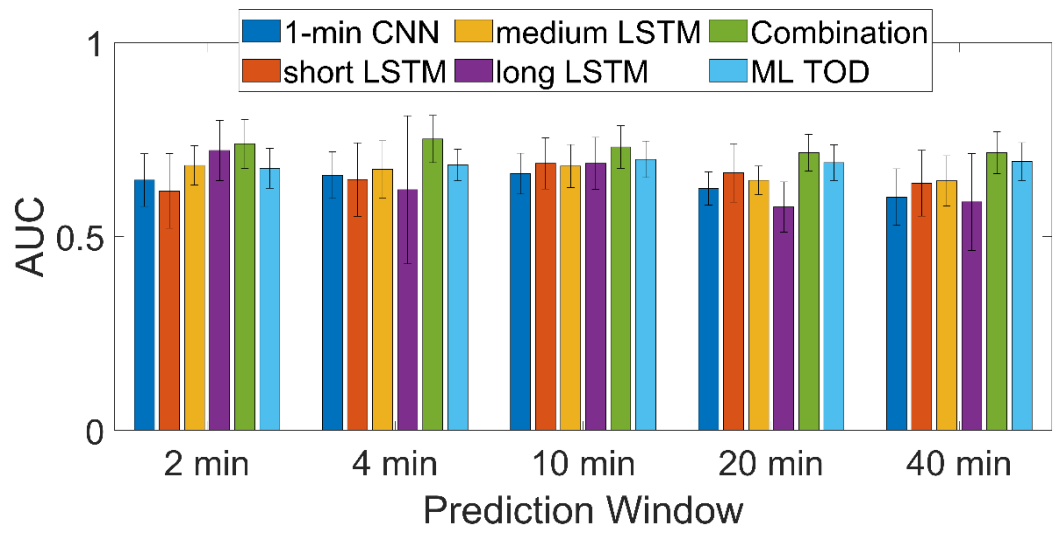